\documentclass[journal]{IEEEtran}
\usepackage{cite}
\usepackage[pdftex]{graphicx}
\usepackage[pdftex]{color}
\graphicspath{/images/}
\usepackage[ruled,vlined,linesnumbered]{algorithm2e}
\usepackage[cmex10]{amsmath}
\usepackage{algpseudocode}
\usepackage{array}
\usepackage{url}
\usepackage{mathtools}
\usepackage{color}
\usepackage{amsmath, amssymb}

\usepackage{amsthm}

\theoremstyle{remark}

\theoremstyle{definition}

\begin{document}
\title{Analysis of the LTE Access Reservation Protocol for Real-Time Traffic}

\author{\IEEEauthorblockN{Henning Thomsen, Nuno K. Pratas, \v Cedomir Stefanovi\' c, Petar Popovski}
\thanks{The authors are with the Department of Electronic Systems, Aalborg University, Aalborg, Denmark (e-mail: \{ht, nup, cs, petarp\}@es.aau.dk).}
\thanks{The research presented in this paper was partly supported by the Danish Council for Independent Research (Det Frie Forskningsr{\aa}d) within the Sapere Aude Research Leader program, Grant No. 11-105159 ``Dependable Wireless Bits for Machine-to-Machine (M2M) Communications'' and performed partly in the framework of the FP7 project ICT-317669 METIS, which is partly funded by the European Union. The authors would like to acknowledge the contributions of their colleagues in METIS, although the views expressed are those of the authors and do not necessarily represent the project.}}

\maketitle

\begin{abstract}
LTE is increasingly seen as a system for serving real-time Machine-to-Machine (M2M) communication needs.
The asynchronous M2M user access in LTE is obtained through a two-phase access reservation protocol (\emph{contention} and \emph{data} phase).
Existing analysis related to these protocols is based on the following assumptions: 
(1) there are sufficient resources in the \emph{data phase} for all detected contention tokens, and (2) the base station is able to detect collisions, i.e., tokens activated by multiple users.
These assumptions are not always applicable to LTE - specifically, (1) due to the variable amount of available data resources caused by variable load, and (2) detection of collisions in contention phase may not be possible.
All of this affects transmission of real-time M2M traffic, where data packets have to be sent within a deadline and may have only one contention opportunity.
We analyze the features of the two-phase LTE reservation protocol and asses its performance, when assumptions (1) and (2) do not hold.
\end{abstract}


\IEEEpeerreviewmaketitle

\section{Introduction}
\label{sec:Introduction}

An access reservation protocol is instrumental in any multi-user communication system in order to enable users to connect asynchronously or transmit intermittently~\cite{Bertsekas1987}.
The Long Term Evolution (LTE) system~\cite{3GPPTS36.201} uses an access protocol consisting of two phases: a \emph{contention phase}, where each user contends by activating a particular reservation token chosen from the set of available tokens; and a \emph{data phase}, where the reservation tokens (i.e., token holders) detected by the base station (BS) get assigned resources for the data transfer.
The asynchronous access in LTE gains importance as the needs to support traffic related to Machine-to-Machine (M2M) communications gets increasingly important.
In many cases, M2M traffic is a real-time traffic, where data packets become obsolete after a deadline and thus may undergo only a single contention and data phases, i.e., unsuccessful transmissions cannot be postponed for later contention or scheduled to a later data phase.

The available analysis of the two-phase access reservation protocols typically assumes that: (1) there are sufficient resources in the data phase to serve all detected reservation tokens; (2) the BS is able to discern between reservation tokens activated by one or more than one users, i.e., the contention phase has a ternary output (idle, single or collision).
However, assumption (1) does not hold in cellular networks such as LTE, where the data phase has limited number of resources, while the network load is variable; this implies that there is a possibility that the users with real-time traffic that contended successfully may not be assigned a data transmission slot at all~\cite{3GPPTS36.201}.
Assumption (2), by default does not hold in LTE, as the BS may not able to discern if a token was activated by one or multiple users~\cite[Sec. 17.5.2.3]{sesia2011lte}.
In other words, there are practical setups in which the BS can ``see'' that a preamble has been activated, but it does not know how many users activated it.
This implies that in the contention phase, collisions ``over'' tokens are treated as singles, i.e., the output of the contention phase is binary (idle or active) instead of the commonly assumed ternary output (idle, single or collision).

In LTE, a reservation token is activated by transmitting a specific preamble in the random access sub-frame~\cite{3GPPTS36.321}.
The preambles are chosen from the orthogonal set of preambles obtained from Zadoff-Chu sequences~\cite{3GPPTS36.213}.
Due to orthogonality of the preambles, the LTE contention phase can be modeled as a framed slotted ALOHA scheme, where ``slots'' represent preambles over which the users contend.

Asynchronous M2M communications based on LTE are considered in several works found in the literature. In~\cite{jiang2012fast}, the authors propose a preamble retransmission method, subject to optimization of the transmission rate. In~\cite{EURECOM+3938}, a packet aggregation method is proposed. Here, M2M devices do not necessarily transmit their packets immediately, but buffer them until a certain threshold. A closed form expression of the collision probability of M2M traffic at the LTE contention phase is provided in~\cite{cheng2012rach}. A recent work using a combinatorial model to study the random access in LTE is given in~\cite{6211364}. Different from~\cite{6211364}, we also consider the data phase in our analysis.

An early study of the ALOHA protocol in a reservation framework was done in~\cite{crowther1973system}, and~\cite{szpankowski1983analysis} consideres two reservation methods based on framed ALOHA.
Access reservation protocols with several parallel data channels have been studied in~\cite{han2006analyzing}. Herein, the authors find the optimal ratio of control to data channels, and the optimal number of data channels, in terms of throughput. Note that in contrast to~\cite{han2006analyzing}, in this letter, we consider collisions in the contention phase to be non-destructive.
Finally, we point out a recent analysis of random access protocols in the context of RFID systems given in \cite{alcaraz2013stochastic}; the results presented therein are directly applicable to the contention phase of the access reservation protocol and are used as a starting point for the analysis presented in the paper.

In this letter, we derive the exact probability mass function (pmf) that a number of reservation tokens activated by a single user are assigned resources in the data phase, when assumptions (1) and (2) do not hold.
Based on the obtained results, we calculate the corresponding one-shot success rate and efficiency of the LTE access reservation scheme.
The presented analysis and derived results are directly applicable to the LTE access reservation scheme for the increasingly important case of asynchronously served users with real-time constraints.

The remainder of this paper is organized as follows.
In Section~\ref{sec:SystemModel}, we present the system model.
Section~\ref{sec:AnalyticalModel} elaborates the method to obtain the exact pmf of the number of reservation tokens that are activated by a single user and that are assigned resources in the data phase, following by the derivation of the success rate and efficiency.
Examples demonstrating derived results are given in Section~\ref{sec:NumericalExamples}, while the letter is concluded in Section~\ref{sec:Conclusion}.

\section{System Model}
\label{sec:SystemModel}
\begin{figure}
	\centering
		\includegraphics[width=\columnwidth]{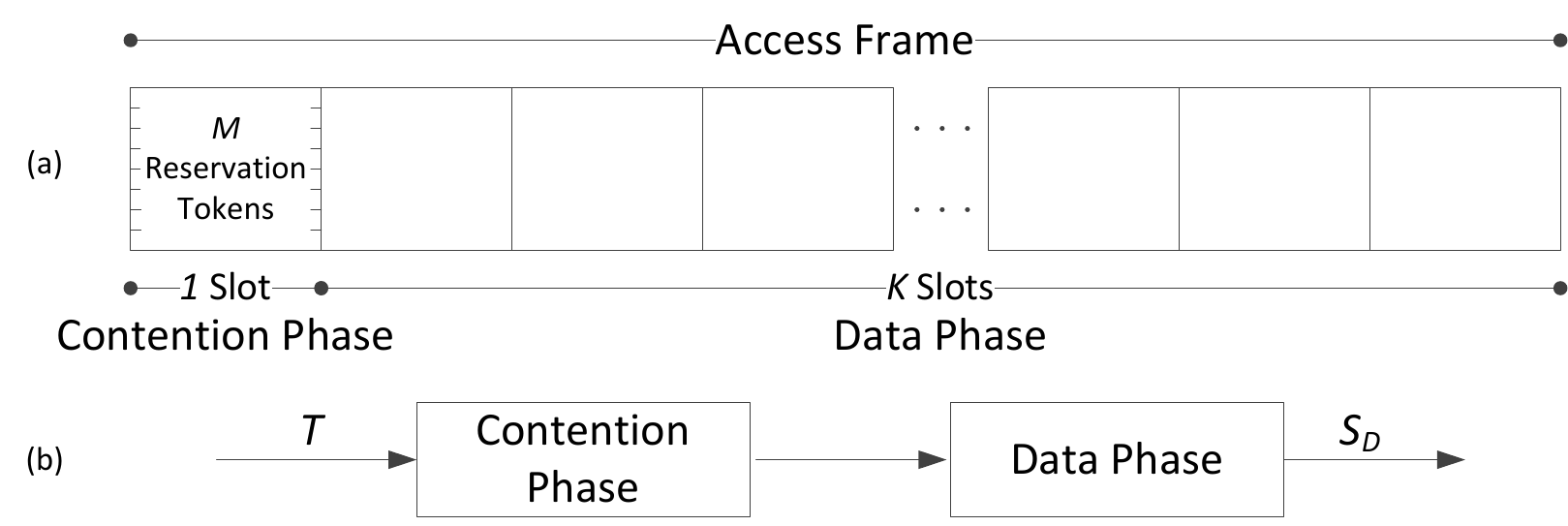}
	\caption{(a) Access Frame and (b) System Model.}
	\label{fig:SystemModelReduced}
\end{figure}

Fig.~\ref{fig:SystemModelReduced}(a) shows a simplified version of the LTE access reservation protocol that captures the details essential for the presented analysis~\cite{3GPPTS36.321}. 
The access reservation is composed of a contention phase and data phase.
The contention phase lasts a single slot and is modeled as a variant of a framed slotted ALOHA, where users contend over a set of of available tokens\footnote{In contrast to standard ALOHA, in the considered model there are no collisions, as elaborated in the letter.}; we assume that in this slot, there are available $M$ reservation tokens for contention.
The data phase is a actually a Time Division Multiplexing (TDM) scheme and we assume that there are available $K$ resource slots (i.e., $K$ TDM slots).
The combination of the two phases is denoted as an access frame, consisting of $K+1$ slots in total.
Finally, we assume that there are $T$ users contending for the available resource slots.

The access reservation protocol operates as follows:
\begin{enumerate}
	\item Each of $T$ users activates randomly and independently one of the $M$ available tokens. A token can be activated by more than one user.
	\item Base Station (BS) detects all activated tokens, irrespective whether they have been activated by one or several users~\cite[Sec. 17.5.2.3]{sesia2011lte}. The BS chooses uniformly randomly $K$ tokens from the set of detected tokens.
	\item The selected tokens are assigned a resource slot each and the corresponding users, i.e., token holders, are informed about the respectively assigned slots through the feedback channel.
	\item The selected token holders transmit their data packets in the assigned resource slots. If two or more holders activated the same token and thus were assigned the same resource slot, their transmissions collide and are considered as lost.
\end{enumerate}

The assumption that the BS is unable to detect collision in contention phase holds in small cells~\cite[Sec. 17.5.2.3]{sesia2011lte}, and refers to the worst case scenario where the detected preamble does not reveal anything about the number of users that transmitted it\footnote{Note that, in practical LTE systems, if the cell size is more than twice the distance corresponding to the maximum delay spread, the BS may, in some circumstances, be able to differentiate the transmission of the same preamble by two or more users, provided that the users are separable in terms of the Power Delay Profile \cite{sesia2011lte}. However, the analysis of such operation is straightforward and therefore not of interest in this letter.}.
Obviously, if the BS knows that there are two or more users using a certain preamble, then a straightforward way to operate is not to assign any resource slot to the preamble, thus preventing collisions and the respective resource waste in the data phase.

\section{Analysis}
\label{sec:AnalyticalModel}

\begin{table}[t]
	\centering
			\begin{tabular}{ l  l }
		  	\hline
  			Variable & Description\\
				\hline
				$M$ & No. of reservation tokens \\
				$K$ & No. of TDM slots \\
				$T$ & No. of accessing users \\
				$S_{D}$ & No. of users succeeding in the data phase \\
				$s$ & No. of reservation tokens selected by one user \\
				$c$ & No. of reservation tokens selected by more than one user \\
				$k$ & $\min\{s+c,K\}$ \\
				$m$ & $\min\{M,T\}$ \\
				$\sigma$ & Success Rate \\
				$\rho$ & Efficiency \\
  			\hline
			\end{tabular}
	\caption{Definition of used variables.}
	\label{tab:DescriptionOfTheRACHModelValues}
\end{table}
\subsection{Derivation of the pmf}
\label{sec:DerivationOfThePmf}
The conditional probability of having assigned $S_{D}=d$ contention tokens used exactly by one user each from the contention phase of size $M$ to the data phase of size $K$ when $T = t$, denoted by $P(S_D = d\mid T = t, K, M)$, is derived in this section.
The parameters used in the derivation are listed in Table~\ref{tab:DescriptionOfTheRACHModelValues}.
\begin{figure*}[!t]
\normalsize
\begin{equation}
\label{eqn:pmftwocol}
P(S_D = d \mid T = t, K, M) = \sum_{0 \leq s \leq m} \sum_{0 \leq c \leq m - s} \frac{\binom{M}{s}T\cdot(T-1)\cdots(T-s+1)\binom{M-s}{c}S_2(T-s,c)c!}{M^T} \frac{\binom{s}{d}\binom{c}{k-d}}{\binom{s+c}{k}}
\end{equation}
\hrulefill
\vspace*{4pt}
\end{figure*}

Suppose that out of the $M$ available tokens, $s$ are used by exactly one user each (singles), and $c$ are used by two or more users (collisions), where $0 \le s+c \le M$.
From the $M$ available tokens, $s$ single tokens can be selected in $\binom{M}{s}$ ways.\footnote{We define $\binom{n}{k} = 0$ when $n < k$.}
Further, as the tokens are distinguishable, the number of ways in which $s$ of the $T$ users can be selected is $T \cdot (T-1) \cdots (T-s+1)$ ways.

From the remaining $M-s$ tokens, we choose $c$ for the colliding users, which can be done in $\binom{M-s}{c}$ ways.
As the tokens are distinguishable, we need to count all permutations of them, which equals $c!$.
Further, the number of ways in which $T-s$ users can be distributed among $c$ tokens such that there are at least two users selecting each token is given
by the 2-associated Stirling number of the second kind $S_2(T-s,c)$ \cite[pp.221-222]{comtet1974advanced}, which can be computed using the recurrence relation~\eqref{eqn:stirlingrecurrence} exposed in the Appendix.

The total number of ways in which $T$ users can select among $M$ tokens without restriction is $M^T$. Therefore, the probability of the $T$ users selecting among $M$ tokens such that there are $s$ tokens used by exactly one user each, and $c$ tokens used by two or more users each, equals
\begin{equation}
	\frac{\binom{M}{s}T\cdot(T-1)\cdots(T-s+1)\binom{M-s}{c}S_2(T-s,c)c!}{M^T},
\end{equation}
as similarly derived in~\cite{alcaraz2013stochastic}, although in a different context.

In the data phase, the $s+c$ used tokens are mapped randomly to the $K$ resource slots. Here we distinguish between two cases, $s+c \le K$ and $s+c > K$. In the first case, all used tokens are assigned to data slots. In the second case, $K$ out of the $s+c$ tokens are randomly selected and assigned to the data slots. We then have that the probability of selecting $d$ slots out of the $K$, such that the selected slots are used by one user each, is given by the hypergeometric distribution
\begin{equation}\label{eqn:hypergeometric}
	\frac{\binom{s}{d}\binom{c}{k-d}}{\binom{s+c}{k}}.
\end{equation}
where $k = \min\{s+c,K\}$. Note that when $k = s+c$, then Eq.~\eqref{eqn:hypergeometric} equals $1$ when $s=d$, and $0$ otherwise.

The probability of selecting $d$ slots containing one user each, given that $T=t$ users transmit, is obtained by summing over all cases of $s$ and $c$ such that $0 \leq s+c \leq M$, which yields the complete expression of $P(S_D = d \mid T = t, K, M)$ presented in Eq.~\eqref{eqn:pmftwocol}, where $m = \min\{M,T\}$ and $k = \min\{s+c,K\}$.

We conclude by noting that \eqref{eqn:pmftwocol} holds when $M>K$ and also when $M \leq K$.

\subsection{Success Rate and Efficiency}
\label{sec:MeanSystemThroughput}

We use the pmf derived in the previous subsection in obtaining an expression for the success rate, defined as the expected value of $S_{D}$ given a number of transmitting users $T$
\begin{equation}
\label{eq:throughput}
	\sigma = \frac{E\left[ S_{D} \vert T = t , K, M\right]}{t} = \frac{1}{t} \sum_{s=0}^{\min\{K,M\}} s \cdot P(S_{D} = s \vert T = t).
\end{equation}
As defined, the success rate takes into account both phases of the LTE access reservation scheme and measures the expected fraction of successfully accessing users.

In order to assess how well the slots of the access frame (see Fig.~\ref{fig:SystemModelReduced}(a)) are utilized, we define the efficiency, given $T=t$, $K$ and $M$, as
\begin{equation}\label{eqn:efficiency}
	\rho = \frac{E\left[ S_{D} \vert T = t , K, M\right]}{K+1}
\end{equation}
where the denominator corresponds to the length of the access frame in slots.

\section{Results}
\label{sec:NumericalExamples}
In this section, we give examples of the pmf in Eq.(\ref{eqn:pmftwocol}), the success rate in Eq.(\ref{eq:throughput}) and efficiency in Eq.(\ref{eqn:efficiency}), both for the cases when $M>K$ and $M \leq K$.

\subsection{pmf Evaluation}
\label{sec:pmfEvaluation}
We first describe the method to obtain the pmf from the simulation. The method can be summarized in the following three steps.
\begin{enumerate}
	\item Let the $T$ users select their tokens randomly and independently. For every token, count the number of times this token is selected by the users.
	\item To simulate a limited data phase, select a random subset of size $K$ from the set of tokens. Count the number of successful tokens in this subset, i.e. tokens used by one user. Call this number $S$.
	\item Let $S^{(n)}$ be the value of $S$ obtained in iteration $1 \leq n \leq N$, where $N$ is the total number of iterations. The simulated pmf is then
	\begin{equation}
		\widehat{P}(S_D = s \mid T = t) = \frac{\sum_{n=1}^{N} \mathbb{I}(S^{(n)})}{N},
	\end{equation}
	where $\mathbb{I}( \cdot )$ is an indicator function, giving $\mathbb{I}(S^{(n)}) = s$ if $S^{(n)} = s$, and $0$ otherwise.
\end{enumerate}

\begin{figure}[t]
	\centering
		\includegraphics[width=\columnwidth]{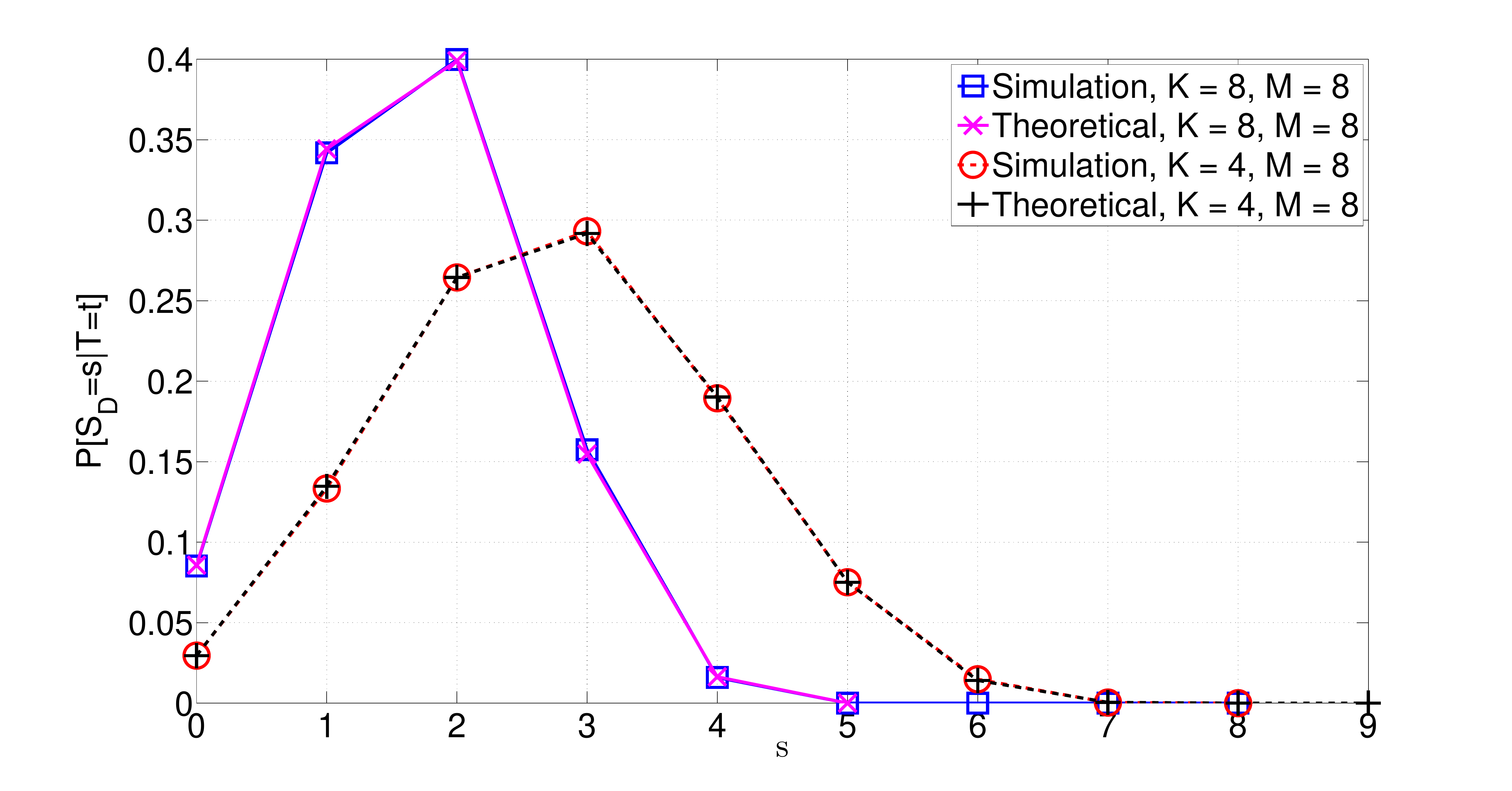}
	\caption{Comparison of theoretical and simulated pmfs for $M>K$ and $M \leq K$.}
	\label{fig:ProbDistCompare}
\end{figure}

\begin{figure*}[t]
\begin{minipage}[b]{0.46\linewidth}
	\centering
		\includegraphics[width=\columnwidth]{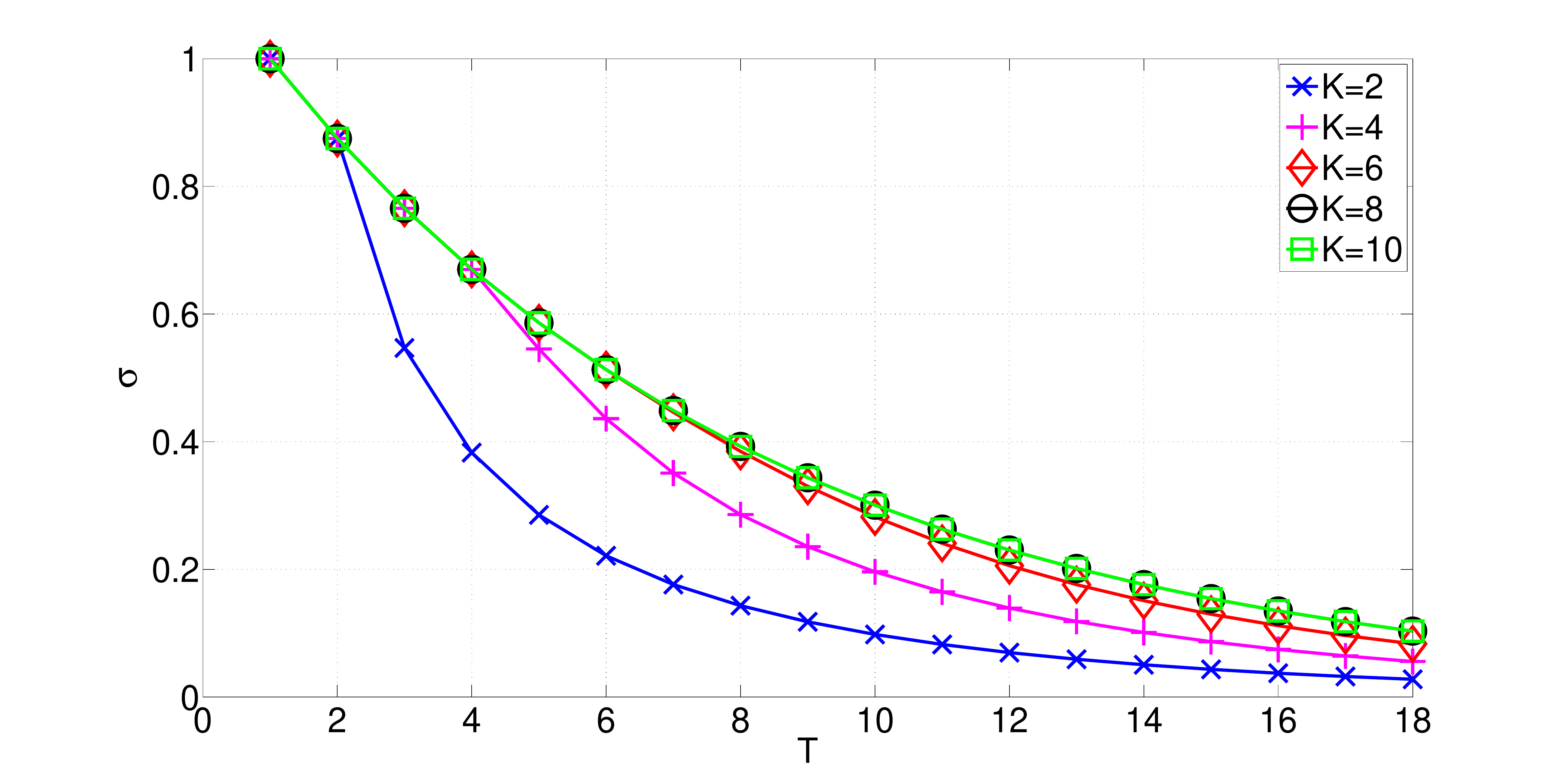}
	\caption{Comparison of success rate $\sigma$, for $M=8$, varying user load $T$, and different sizes of the data phase $K$.}
	\label{fig:ExpectedNumberOfSuccessesCompare}\end{minipage}
\hspace{0.7cm}
\begin{minipage}[b]{0.46\linewidth}
	\centering
		\includegraphics[width=\columnwidth]{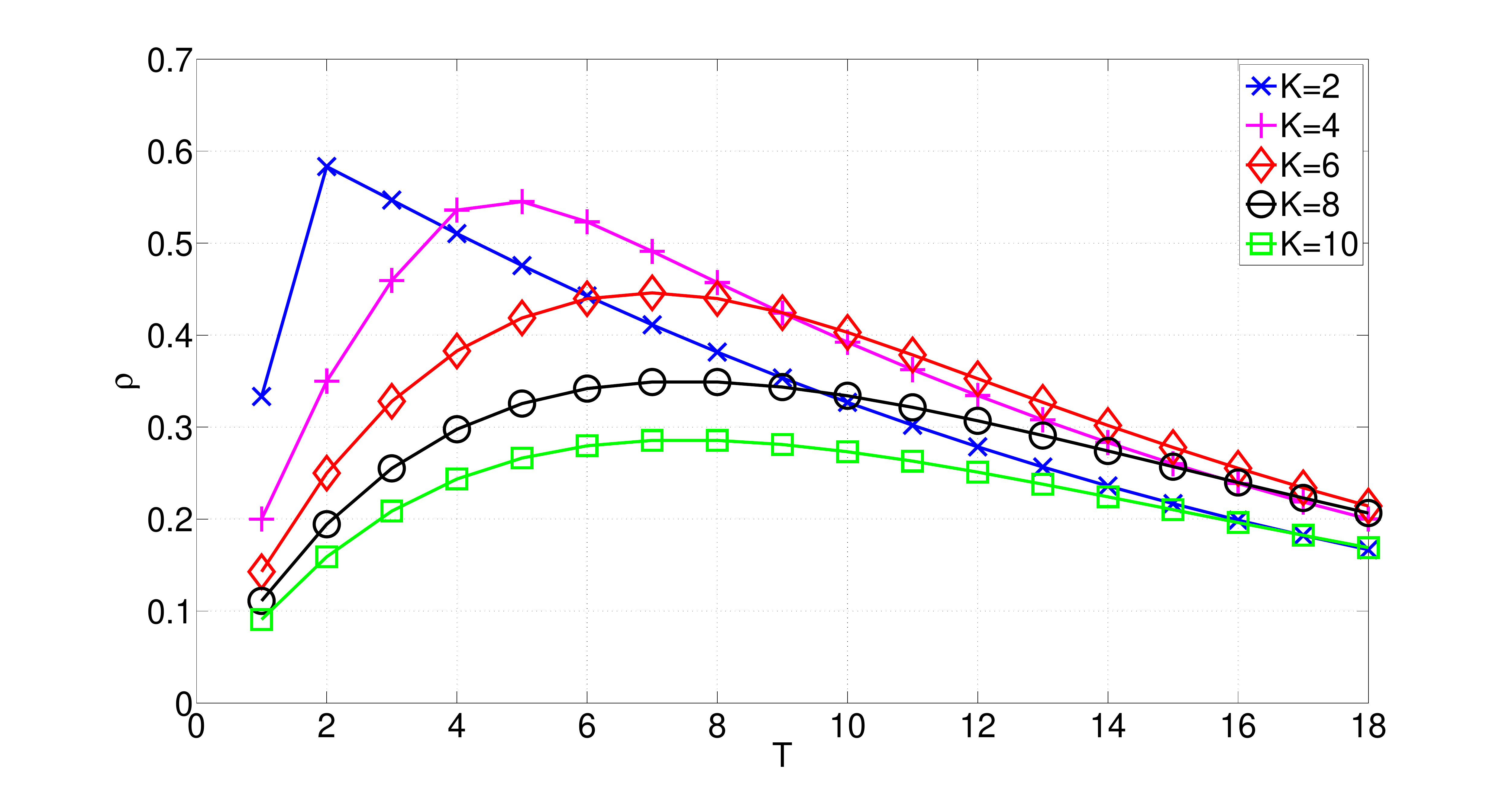}
	\caption{Comparison of efficiency $\rho$, for $M=8$, varying user load $T$, and different sizes of the data phase $K$.}
	\label{fig:EfficiencyCompare}\end{minipage}
\end{figure*}

Fig.~\ref{fig:ProbDistCompare} compares the analytical and simulated pmfs for $T=12$ accessing users, when number of tokens is set to $M=8$ and the number of resource slots is $K=4$ and $K=8$.
The analytical pmf is obtained from Eq.~\eqref{eqn:pmftwocol}, while the simulated one is obtained from running $N = 100000$ simulation iterations.
We observe a correlation between the analytical and simulated pmfs, validating the presented analysis.

\subsection{Success Rate and Efficiency Evaluation}
\label{sec:SuccessRateAndEfficiencyEvaluation}

Fig.~\ref{fig:ExpectedNumberOfSuccessesCompare} and Fig.~\ref{fig:EfficiencyCompare} show respectively the success rate $\sigma$, Eq.~\eqref{eq:throughput}, and the efficiency $\rho$, Eq.~\eqref{eqn:efficiency} as functions of number of users $T$, for $M=8$ and varying $K$.

From Fig.~\ref{fig:ExpectedNumberOfSuccessesCompare} it can be observed that success rate decreases as the number of users $T$ increase; this is due to increasing probability of multiple users selecting the same tokens and, consequently, increasing number of collisions happening in the data phase.
On the other hand, increasing the number of available resource slots $K$ increases the success rate at first, until $K$ reaches the number of available tokens $M$. Afterwards, there is no benefit in increasing $K$, as no more than $M$ users can be successfully detected and admitted in the system and $K-M$ resource slots will always be left unassigned.

From Fig.~\ref{fig:EfficiencyCompare} it is clear that the higher efficiency is achieved when is $K$ considerably lower than $M$, which is not a straightforward conclusion.
Again, this is because of multiple users selecting the same preambles, which results both in \emph{collision} and \emph{idle} slots in the data phase; the chances for the latter increase with the number of available resource slots $K$. 
We end this section by noting that, by using the framework presented in paper, the optimal number of resource slots $K$ that maximizes the efficiency can be calculated, for fixed number of tokens $M$ and number of users $T$.

\section{Conclusion}
\label{sec:Conclusion}

In this letter we studied a LTE based access reservation protocol and provided a method to obtain the exact pmf describing the number of reservation tokens activated by a single user that gets assigned resources in the data phase.
The obtained results are applicable to the case where there are not enough resources in the data phase to serve all detected reservation tokens, and when the base station is not able to discern between reservation tokens selected by one or more than one user; both assumptions may occur in practice, affecting the operation of the access reservation protocol.

Further, based on the presented method we derived the one-shot success rate and efficiency of the scheme, which can be used as performance measures of the constrained access reservation systems, such as LTE, in the emerging scenarios with real-time M2M communication, when there is a limited time to carry out the contention and the data transmissions.
Finally, it was observed that although the success rate is maximized when there are the same number of contention and data resources, there is a non-negligible efficiency tradeoff for doing so.

\appendix
Here, a method to compute the 2-associated Stirling numbers of the second kind is presented, based on~\cite{comtet1974advanced}.
By definition, $S_2(n,k)$ is the number of ways in which $k$ objects can be put into $n$ boxes, such that each box contains at least $2$ objects.
The values of $S_2(n,k)$ can be computed using the recurrence relation
\begin{equation}\label{eqn:stirlingrecurrence}
	S_2(n+1,k) = k S_2(n,k) + n S_2(n-1,k-1),
\end{equation}
with initial condition $S_2(2,1) = 1$. If $k > \lfloor \frac{n}{2} \rfloor$, $n \le 0$ or $k \le 0$, then $S_2(n,k) = 0$.


\end{document}